\documentclass[reprint,amsmath,amssymb,aps,prb]{revtex4-2}

\usepackage{adjustbox}
\usepackage{placeins}
\usepackage{newtxtext}
\usepackage[varvw]{newtxmath}
\usepackage{graphicx}
\usepackage{dcolumn}
\usepackage{balance}
\usepackage{bm}
\usepackage{bbm}
\usepackage{csquotes}
\usepackage{mathtools}
\usepackage{booktabs,makecell}
\usepackage{physics}
\usepackage{subfigure}
\usepackage{tikz}
\usepackage{tikz-feynman}
\usetikzlibrary{decorations.pathreplacing}

\usepackage{amsfonts,amsmath,amssymb,amsthm}
\usepackage[colorlinks=true,allcolors=blue,breaklinks=True]{hyperref}

\newcommand{\klr}[1]{\left(#1\right)}
\newcommand{\kle}[1]{\left[#1\right]}
\newcommand{\kls}[1]{\left|#1\right|}

\newcommand{\klsr}[1]{#1 \biggl|}
\newcommand{\kla}[1]{\langle#1\rangle}
\newcommand{\klg}[1]{\left\{#1\right\}}

\newcommand{\varvec}[1]{\mathbf{#1}}
\renewcommand{\vv}[1]{\varvec{#1}}

\newcommand{\svec}[1]{\begin{pmatrix}#1\end{pmatrix}}
\newcommand{\matr}[1]{\svec{#1}}
\newcommand{\varvarvec}[1]{\bm{#1}}
\newcommand{\vvv}[1]{\varvarvec{#1}}

\newcommand{\psiast}[1]{{\psi^\ast}\!\!_{#1}}



\begin{document}

\title{Quantum-geometric thermal conductivity of superconductors}

\author{Maximilian Buthenhoff}
\email{buthenhoff.m.b890@m.isct.ac.jp}
\affiliation{Department of Physics, Institute of Science Tokyo, Ookayama, Meguro, Tokyo 152-8551, Japan}

\author{Yusuke Nishida}
\affiliation{Department of Physics, Institute of Science Tokyo, Ookayama, Meguro, Tokyo 152-8551, Japan}

\date{\today}

\begin{abstract}
By coupling Bardeen-Cooper-Schrieffer (BCS) theory with isolated bands to an external gravitomagnetic vector potential via a gravitomagnetic Peierls substitution, we identify a quantum-geometric contribution to the electronic contribution of the thermal conductivity. This contribution is governed by the quantum metric in the parameter space spanned by the components of the external gravitomagnetic vector potential which corresponds to a weighted quantum metric in momentum space. In the flat-band limit, we establish an upper and lower Wiedemann-Franz-type bound for the ratio of thermal Meissner stiffness and electric Meissner stiffness (superfluid weight), whose prefactors are provided by the extrema of the squared energy offsets of the outer single-particle bands of the system. Similarly to the superfluid weight, this also leads to a lower bound of the thermal Meissner stiffness in terms of the Chern number. Our results apply to both superconductors and other fermionic superfluids.
\end{abstract}

\maketitle

\section{Introduction}
In a series of papers in the 1950s and 1960s~\cite{bardeen1959theory,tewordt1963theory,ambegaokar1964theory,ambegaokar1965theory}, the electronic contribution to the thermal conductivity of $s$-wave superconductors has been analyzed and calculated with the help of the Boltzmann equation and the Kubo formula~\cite{kubo1957statistical}. The treatment for an anisotropic superconducting state has been further generalized by Hirschfeld, W\"olfle, and Einzel in 1987~\cite{hirschfeld1988consequences}, which also served as a basis for the discussion of the thermal conductivity in superconducting $\mathrm{UTe}_2$~\cite{mishra2024thermal,hayes2025robust}, and other high-temperature superconductors~\cite{tewordt1989theory,uher1990thermal}. \\
\indent However, in 2006, Shastry pointed out in Ref.~\cite{shastry2006sum}, that there arises an additional nontrivial correction to the Kubo formula for the electronic thermal conductivity which does not necessarily vanish for nondissipative systems such as superconductors and superfluids \cite{nakai2022energy}
\begin{align}
    \kappa(\omega) = \frac{iD_{\mathrm{Q}}}{T(\omega + i\delta)} + \kappa_{\mathrm{Kubo}}(\omega) \,.
    \label{eq:thermal-cond}
\end{align}
The additional contribution is proportional to the thermal Meissner stiffness $D_{\mathrm{Q}}$. Its physical origin is analogous to that of the superfluid weight (or electric Meissner stiffness), in that the macroscopic phase rigidity gives rise to a phase stiffness against an external vector potential. In analogy to the electric Meissner stiffness, which measures the energy cost to create a modulation of the order parameter phase~\cite{peotta2025quantum}, the thermal Meissner stiffness quantifies the energy needed to thread a static gravitomagnetic flux that induces a persistent heat current~\cite{nakai2022energy,Nakai2023twisted}. Equivalently, while the superfluid weight can be viewed as the stiffness against an electromagnetic phase twist, the thermal Meissner stiffness is the stiffness against a gravitomagnetic vector potential \cite{mashhoon2007measurement,furusaki2013electromagnetic,golkar2016global,nakai2017laughlin} that couples to heat current. Specifically, a static uniform gravitomagnetic field is equivalent to describing a system in a uniformly rotating frame with constant angular velocity~\cite{furusaki2013electromagnetic}. \\
\indent Related gravitoelectromagnetic effects have been thematized previously. For example, in Refs.~\cite{Tajmar2003gravitational,Graham2008experiment} it is discussed whether gravitoelectromagnetic effects may explain discrepancies found in tests of superconducting gyroscopes \cite{Gallerati2022interaction}, and in Ref.~\cite{Sekine2016chiral} a thermal counterpart of the chiral magnetic effect~\cite{fukushima2008chiral}, the chiral gravitomagnetic effect (a thermal current generated by gravitomagnetic fields) is studied in superconductors and superfluids. \\
\indent The goal of this work is to calculate the thermal Meissner stiffness for fermionic superfluids (such as superconductors) with isolated bands, described by Bardeen-Cooper-Schrieffer (BCS) theory~\cite{bardeen1957theory}, that hosts a pairing mechanism with a time-reversal-symmetry-preserving gap function. Under these assumptions, the thermal Meissner stiffness $D_{\mathrm{Q}}$ can be expressed in terms of the second-order partial derivative of the grand potential $\Omega$ with respect to the components $\lambda_i$ of the gravitomagnetic vector potential 
\begin{align}
    D_{\mathrm{Q},ij} = \frac{1}{V} \frac{\partial^2 \Omega}{\partial \lambda_i \partial \lambda_j} \klsr{}_{\vvv{\uplambda} = 0} \,;
    \label{eq:thermal-meissner-here}
\end{align}
see Appendix~\ref{appendix} for more details. Using this formula, we identify two contributions to the thermal Meissner stiffness, one of them being driven by the dispersion of the single-particle bands. The other is determined by the quantum metric in the parameter space spanned by the components of the external gravitomagnetic vector potential, and admits lower and upper bounds in terms of the usual quantum metric in momentum space. This quantum-geometric quantity is referred to as the \enquote{heat quantum metric} in Ref.~\cite{bermond2025dichroism} and as the \enquote{thermal quantum metric} in Ref.~\cite{lhachemi2026unifying}, and a possible experimental measurement of it is proposed. The quantum metric is part of the quantum geometry of a quantum material and represents a gauge invariant measure for infinitesimal distances of the Hilbert-Schmidt quantum distance \cite{yu2025quantum}, and thus quantifies the overlap of eigenstates at nearby points in parameter space \cite{provost1980riemannian,berry1989quantum}. In analogy to the Wiedemann-Franz law, which states that the ratio of electrical to thermal conductivity is set by a universal coefficient that is linear in temperature \cite{franz1853ueber,smrcka1977transport,graf1996electronic}, we show that, in the limit of an isolated flat band, the ratio between the thermal Meissner stiffness and the superfluid weight follows a Wiedemann-Franz-type inequality. The bounds are material dependent and set by the dispersions of the outer bands of the system.
\section{Gravitomagnetic Peierls substitution}
To introduce the external gravitomagnetic vector potential $\vvv{\uplambda}$ in the single-particle Hamiltonian to analyze thermal properties of the system, we follow the strategy of Refs.~\cite{golkar2016global,nakai2022energy}. The construction is related to that of Refs.~\cite{luttinger1964theory,furusaki2013electromagnetic,tatara2015thermal,sekine2020quantum,amitani2024universal} where an external field conjugate to the energy density is introduced. Here, the gravitomagnetic vector potential serves merely as a theoretical tool to calculate the thermal Meissner stiffness analytically. Consider a $(d+1)$-dimensional system on a torus parametrized by $(\tau,\vv{x})$ with metric
%
    $\mathrm{d} s^2 = (\mathrm{d}\tau + i\lambda_i \mathrm{d}x^i)^2 + \delta_{ij} \mathrm{d}x^i \mathrm{d}x^j$,
%
where the vector field $\lambda_i$ is the background field called graviphoton or gravitomagnetic vector potential. We impose that $\tau$ has periodicity $\beta$ and $x_i$ has periodicity $L_i$. Accordingly, in the system with metric defined above we have the identifications $(\tau,\vv{x}) \sim (\tau + \beta,\vv{x}) \sim (\tau + i\lambda_i L_i, \vv{x} + L_i \vv{e}_i)$, where $\vv{e}_i$ is the $i$th unit vector. \\
\indent Consider the $n$th eigenstate of the single-particle Hamiltonian $H(\vv{k})$ with eigenvalue $\varepsilon_n(\vv{k})$, $n = 1,\hdots,N_{\mathrm{B}}$, and denote by $\xi_n(\vv{k}) = \varepsilon_n(\vv{k}) - \mu$ the energy offsets of the energy bands from the chemical potential. By applying Bloch's theorem to the grand-canonical single-particle Hamiltonian $K(\vv{k}) \coloneqq H(\vv{k}) - \mu\mathbbm{1}$, the twisted boundary condition above implies the quantization condition $(k_i - \xi_n(\vv{k})\lambda_i) L_i = 2\pi n_i$ with $n_i \in \mathbb{Z}$.
%
%
At $\vvv{\uplambda} = 0$, the allowed momenta are $ k_i = {2\pi n_i}/{L_i}$ and the corresponding energies are given by $\xi_n(\vv{k})$. On the other hand, at $\vvv{\uplambda} \neq 0$, the same integer $n_i$ labels a state with twisted momentum $\vv{k}^{(\vvv{\uplambda})}(\vv{k})$ satisfying the implicit equation $\vv{k}^{(\vvv{\uplambda})}(\vv{k}) = \vv{k} + \xi_n(\vv{k}^{(\vvv{\uplambda})}(\vv{k})) \vvv{\uplambda}$. Hence, for each band we observe a twist in the single-particle energies. In particular, if we define the twisted energies by $\xi_n^{(\vvv{\uplambda})}(\vv{k}) = \xi_n(\vv{k}^{(\vvv{\uplambda})}(\vv{k}))$ for each band, we find a gravitomagnetic analog of the usual Peierls substitution \cite{peierls1933theory} defined via the implicit dispersion relation 
\begin{align}
    \xi_n^{(\vvv{\uplambda})}(\vv{k}) =  \xi_n(\vv{k} + \xi_n^{(\vvv{\uplambda})}(\vv{k})\vvv{\uplambda}) \,, \label{eq:gravitomagnetic-peierls}
\end{align}
called \enquote{gravitomagnetic Peierls substitution.} For single-band systems with constant eigenenergy $\varepsilon$, Eq.~\eqref{eq:gravitomagnetic-peierls} is equivalent to a shift in momentum $\vv{k} \to \vv{k} + \xi \vvv{\uplambda}$, cf.\ Refs.~\cite{nakai2022energy,sekine2020quantum}. If the energy bands are sufficiently smooth in momentum and $\vvv{\uplambda}$ space, Eq.~\eqref{eq:gravitomagnetic-peierls} is equivalent to a inviscid Burgers equation; see also Ref.~\cite{nakai2022energy} for more details. Under assumption of sufficient smoothness, we can find the Taylor expansion of the twisted energies with respect to $\vvv{\uplambda}$ via implicit differentiation. We obtain up to second order 
\begin{align}
    \xi^{(\vvv{\uplambda})}_n(\vv{k}) &= \xi_n(\vv{k}) + \xi_n(\vv{k}) \partial_{k_i}\xi_n(\vv{k}) \lambda_i + \bigg[\frac{1}{2} \xi_n^2(\vv{k}) \partial_{k_i}\partial_{k_j} \xi_n(\vv{k}) \nonumber \\
    &+  \xi_n(\vv{k}) \partial_{k_i} \xi_n(\vv{k})\partial_{k_j} \xi_n(\vv{k})\bigg] \lambda_i\lambda_j + \mathcal{O}(\lambda^3) \,.
    \label{eq:taylor-twisted-energies}
\end{align}
This expansion is useful in the later sections. \\
%
%
\indent In computing the thermal Meissner stiffness for BCS superconductors and, more generally, fermionic superfluids, it is not sufficient to know solely the impact of the external gravitomagnetic vector potential on the single-particle energies. Additionally, we need the Wilczek-Zee connection \cite{wilczek1984appearance} (or nonadiabatic coupling~\cite{yarkony2002nonadiabatic}) in the parameter space spanned by the components of the external gravitomagnetic vector potential, $e_{i,m}^{\ket{\psi_n(\vv{k},\cdot)}}(\vvv{\uplambda}) = \braket{\psi_m(\vv{k},\vvv{\uplambda})}{\partial_{\lambda_i} \psi_n(\vv{k},\vvv{\uplambda})}$, between two distinct states ($n \neq m$) evaluated at $\vvv{\uplambda}=0$. A nonzero Wilczek-Zee connection can be interpreted as the inability of a state to remain in the same state after a small variation of parameters~\cite{romero2024n}, and enters the quantum metric in $\vvv{\uplambda}$ space as
\begin{align}
    g_{ij}^{\ket{\psi_n(\vv{k},\cdot)}}(\vvv{\uplambda}) = \sum_{m \neq n} \mathrm{Re}\kle{\bar{e}_{i,m}^{\ket{\psi_n(\vv{k},\cdot)}}(\vvv{\uplambda}) e_{j,m}^{\ket{\psi_n(\vv{k},\cdot)}}(\vvv{\uplambda})} \,.
    \label{eq:quantum-metric-boost-deformation-space}
\end{align}
Hence, it represents another quantity associated to the quantum geometry of a quantum system. \\
\indent The behavior of the Wilczek-Zee connection in $\vvv{\uplambda}$ space is determined by the heat current operator of the single-particle Hamiltonian, which enters the deformed single-particle Hamiltonian (with convention $\hbar = 1$) according to $K(\vv{k},\vvv{\uplambda}) = K(\vv{k}) - J_i^{\mathrm{Q}}(\vv{k})\lambda^i + \mathcal{O}(\lambda^2)$~\cite{nakai2022energy}. For a multiband single-particle Hamiltonian, the heat current operator has been calculated to be given by $J_i^{\mathrm{Q}}(\vv{k}) = \frac{1}{2} \klg{K(\vv{k}),\partial_{k_i}K(\vv{k})}$~\cite{shastry2006sum,sekine2020quantum,onishi2022theory}.
%
%
Consequently, we find at zero external gravitomagnetic vector potential
\begin{align}
    e_{i,m}^{\ket{\psi_n(\vv{k},\cdot)}}(\vvv{\uplambda} = 0) = -\frac{\xi_n(\vv{k}) + \xi_m(\vv{k})}{2} e_{i,m}^{(n)}(\vv{k})\,,
    \label{eq:relation-boost-deformed-WZ-and-usual-WZ}
\end{align}
where $e_{i,m}^{(n)}(\vv{k}) = \braket{\psi_m(\vv{k})}{\partial_{k_i} \psi_n(\vv{k})}$ is the usual Wilczek-Zee connection in momentum space.
\section{Mean-field Hamiltonian and grand potential}
Denote by $U(\vv{k},\vv{k}')$ an attractive effective two-particle interaction and consider a time-reversal symmetric (TRS) single-particle Hamiltonian $K(\vv{k})$. We introduce an external gravitomagnetic vector potential $\vvv{\uplambda}$ within the single-particle Hamiltonian $K(\vv{k},\vvv{\uplambda})$ by performing the gravitomagnetic Peierls substitution stated in Eq.~\eqref{eq:gravitomagnetic-peierls}. Then, after a Hubbard-Stratonovich transformation and saddle-point approximation, the BCS mean-field Hamiltonian coupled to an external gravitomagnetic vector potential $\vvv{\uplambda}$ is given by
\begin{align}
    H_{\mathrm{MF}}(\vvv{\uplambda}) &= \sum_{\vv{k}} \Phi_{\vv{k}}^\dagger \mathcal{H}_{\mathrm{BdG}}(\vv{k},\vvv{\uplambda}) \Phi_{\vv{k}}
    + \sum_{\vv{k}} \tr K(\vv{k},\vvv{\uplambda}) \nonumber \\
    &+ \frac{V}{2} \sum_{\vv{k},\vv{k}'} U^{-1}(\vv{k},\vv{k}') \Delta^\dagger_{\alpha\beta}(\vv{k}) \Delta_{\beta\alpha}(\vv{k}')\,.
    \label{eq:mf-hamiltonian}
\end{align}
Here, we use the Einstein sum convention in the orbital indices $\alpha = 1,\hdots,N_{\mathrm{B}}$. As in Ref.~\cite{xie2020topology}, we include the spin index in $\alpha$. Further, $V$ represents the volume of the Brillouin zone, $\phi^\dagger_{\vv{k}\alpha}$ and $\phi_{\vv{k}\alpha}$ are creation and annihilation operators of an electron in orbital $\alpha$ with momentum $\vv{k}$, and $\Phi_{\vv{k}} = (\phi_{\vv{k}1},\phi_{\vv{k}2},\hdots,\phi^\dagger_{-\vv{k}1},\phi^\dagger_{-\vv{k}2},\hdots)$ is the Nambu spinor. The Bogoliubov-de Gennes (BdG) Hamiltonian at momentum $\vv{k}$ under presence of an external gravitomagnetic vector potential $\vvv{\uplambda}$ is given by
\begin{align}
    \mathcal{H}_{\mathrm{BdG}}(\vv{k},\vvv{\uplambda}) = \matr{
        K(\vv{k},\vvv{\uplambda}) & \Delta(\vv{k})\\
        \Delta^\dagger(\vv{k}) & -K^T(-\vv{k},\vvv{\uplambda})
    } \,,
    \label{eq:bdg-hamiltonian}
\end{align}
where $\Delta(\vv{k})$ is the TRS preserving gap function determined by a self-consistent equation. \\
\indent In the following we discuss systems with uniform pairing~\cite{peotta2015superfluidity}, i.e., we assume that $\Delta_{\alpha\beta}(\vv{k}) = \Delta(\vv{k})\delta_{\alpha\beta}$ where $\Delta(\vv{k})$ is a real number (for fixed $\vvv{\uplambda}$ and $\vv{k}$). Note that, because the gap function preserves TRS and can be chosen real, we may, similar to Ref.~\cite{huhtinen2022revisiting}, ignore its implicit dependence on the external gravitomagnetic vector potential, even though the gap function inherits its $\vvv{\uplambda}$ dependency from the gravitomagnetic vector potential dependent single-particle Hamiltonian. 
Using TRS, one finds that only derivatives of the grand potential with respect to the imaginary part of the gap function (evaluated at vanishing external gravitomagnetic vector potential) contribute to the thermal Meissner stiffness. However, since we assume the gap functions to be real, the grand potential is independent of these imaginary components. We can therefore safely neglect the dependence of the gap functions on the external gravitomagnetic vector potential. Note that if we consider multiple independent order parameters, we need to work in the minimal quantum metric basis, see Ref.~\cite{huhtinen2022revisiting} for more details. Also, if the gap functions break TRS, we obtain an additional contribution similar to the one discussed in Ref.~\cite{buthenhoff2025functional}. \\
\indent The relevant part of the grand potential of the mean-field Hamiltonian \eqref{eq:mf-hamiltonian} is given by
\begin{align}
    \Omega \supset& -T\sum_{\vv{k},\sigma,n} \ln(1 + \exp\klr{-\frac{E_{\sigma n \vv{k}}(\vvv{\uplambda})}{T}}) + \sum_{\vv{k}} \tr\xi^{(\vvv{\uplambda})}(\vv{k}) \,,
    \label{eq:grand-potential}
\end{align}
where $E_{\pm n\vv{k}}$ are the $2N_{\mathrm{B}}$ eigenvalues of the BdG Hamiltonian, and we recall that the matrix $\xi^{(\vvv{\uplambda})}(\vv{k})$ is defined as the solution of the implicit equation \eqref{eq:gravitomagnetic-peierls}. Denote by $S(\vv{k},\vvv{\uplambda})$ the modal matrix of $K(\vv{k},\vvv{\uplambda})$. Then, the $N_{\mathrm{B}}$ eigenstates of the single-particle Hamiltonian are provided by the columns of the modal matrix according to $\ket{\psi_n(\vv{k},\vvv{\uplambda})} = [S(\vv{k},\vvv{\uplambda})]_{\cdot,n}$. By introducing a set of creation and annihilation operators determined by the diagonalization matrix $S(\vv{k},\vvv{\uplambda})$, the single-particle Hamiltonian becomes diagonal, $\xi^{(\vvv{\uplambda})}(\vv{k}) = S^\dagger(\vv{k},\vvv{\uplambda}) K(\vv{k},\vvv{\uplambda}) S(\vv{k},\vvv{\uplambda})$, while the pairing term is rotated to $\mathcal{D}_{\vv{k}}(\vvv{\uplambda}) = S^\dagger(\vv{k},\vvv{\uplambda})\Delta(\vv{k})S^\ast(-\vv{k},\vvv{\uplambda})$. \\
\indent If the bands are sufficiently isolated, interband couplings can be neglected and the single-particle diagonalized BdG Hamiltonian becomes approximately block diagonal. Therefore, we can decompose it into a direct sum of $(2\times2)$ matrices~\cite{buthenhoff2025functional}
\begin{align}
    \mathcal{H}_{\vv{k}}(\vvv{\uplambda}) \approx \bigoplus_{n=1}^{N_{\mathrm{B}}} \matr{
        \xi^{(\vvv{\uplambda})}_n(\vv{k}) & \mathcal{D}_{n\vv{k}}(\vvv{\uplambda})\\
        \mathcal{D}^\ast_{n\vv{k}}(\vvv{\uplambda}) & -\xi^{(\vvv{\uplambda})}_n(-\vv{k})
    } \,,
    \label{eq:small-bdg-matrix}
\end{align}
where $\mathcal{D}_{n\vv{k}} = \bra{\psi_n(\vv{k},\vvv{\uplambda})}\Delta(\vv{k})\ket{\psiast{n}(-\vv{k},\vvv{\uplambda})}$ are complex numbers, see also the Supplemental Material of Ref.~\cite{xie2020topology}.
%
%
%
By utilizing the expansion given in Eq.~\eqref{eq:taylor-twisted-energies} and making use of TRS, i.e., $\xi_n(\vv{k}) = \xi_n(-\vv{k})$, we obtain the following expression for the quasiparticle energies of the ($2\times2$) matrices building the single-particle diagonalized BdG Hamiltonian:
\begin{align}
    &E_{\pm n\vv{k}}(\vvv{\uplambda}) = \pm \sqrt{\xi_n(\vv{k})^2 + |\mathcal{D}_{n\vv{k}}(\vvv{\uplambda})|^2} + \xi_n(\vv{k}) \partial_{k_i} \xi_n(\vv{k}) \lambda_i \nonumber \\
    &\pm \frac{\xi_n(\vv{k})}{\sqrt{\xi_n(\vv{k})^2 + |\mathcal{D}_{n\vv{k}}(\vvv{\uplambda})|^2}} \bigg[\xi_n(\vv{k})\partial_{k_i}\xi_n(\vv{k}) \partial_{k_j}\xi_n(\vv{k}) \nonumber\\
    &+ \frac{1}{2}\xi_n^2(\vv{k}) \partial_{k_i} \partial_{k_j} \xi_n(\vv{k})\bigg]\lambda_i\lambda_j + \mathcal{O}(\lambda^3) \,.
    \label{eq:quasi-energy}
\end{align}
\section{Quantum-geometric thermal Meissner stiffness}
%
%
According to the chain rule, we find the second derivative of the first term of the grand potential~\eqref{eq:grand-potential} to be
\begin{align}
    &\frac{\partial^2}{\partial\lambda_i \partial\lambda_j} \klr{-T\ln\klr{1 + \exp\klr{-\frac{E_{\pm n\vv{k}}(\vvv{\uplambda})}{T}}}} \nonumber \\
    &= n_{\mathrm{F}}'(E_{\pm n\vv{k}}(\vvv{\uplambda})) \frac{\partial E_{\pm n\vv{k}}(\vvv{\uplambda})}{\partial \lambda_i} \frac{\partial E_{\pm n\vv{k}}(\vvv{\uplambda})}{\partial \lambda_j}\nonumber\\
    &+ n_{\mathrm{F}}(E_{\pm n\vv{k}}(\vvv{\uplambda})) \frac{\partial^2 E_{\pm n\vv{k}}(\vvv{\uplambda})}{\partial\lambda_i\partial\lambda_j} \,.
\end{align}
Therefore, we need to find the first- and second-order derivatives of the quasiparticle energies \eqref{eq:quasi-energy} at vanishing external gravitomagnetic vector potential. \\ 
%
%
\indent Due to the uniform pairing condition the derivative of the offdiagonal elements $|\mathcal{D}_{n\vv{k}}(\vvv{\uplambda})|^2$ of the single-particle diagonalized BdG Hamiltonians is determined by derivatives of the fidelity $F_{n\vv{k}}(\vvv{\uplambda}) = \kls{\braket{\psi_n(\vv{k},\vvv{\uplambda})}{\psiast{n}(-\vv{k},\vvv{\uplambda})}}^2$ between an eigenstate and the complex-conjugated eigenstate at opposite momentum. Since, in the absence of an external gravitomagnetic vector potential, TRS forces the two states to coincide and hence their fidelity to be equal to one, we obtain for the first derivative with respect to $\lambda_i$ at zero external field $\partial_{\lambda_i} F_{n\vv{k}}(\vvv{\uplambda})|_{\vvv{\uplambda} = 0} = 2\,\mathrm{Re}\!\klr{\partial_{\lambda_i} \braket{\psi_n(\vv{k},\vvv{\uplambda})}{\psiast{n}(-\vv{k},\vvv{\uplambda})}|_{\vvv{\uplambda} = 0}}$.
%
%
Denote by $\ket{\zeta_n(\vv{k},\cdot)}$ 
the difference between an eigenstate and the complex-conjugated eigenstate at opposite momentum. By TRS, the overlap of $\partial_{\lambda_i}\ket{\zeta_n(\vv{k},\vvv{\uplambda})}|_{\vvv{\uplambda} = 0}$ with the $n$th eigenstate of the single-particle Hamiltonian at zero external vector potential equals the difference of two Berry connections multiplied by the imaginary number, and is therefore purely imaginary. Consequently, $\partial_{\lambda_i} F_{n\vv{k}}(\vvv{\uplambda})|_{\vvv{\uplambda} = 0} = 0$, such that the first derivative of the energy with respect to $\lambda_i$ at zero external vector potential is given by $\partial_{\lambda_i} E_{\pm n\vv{k}}(\vvv{\uplambda})|_{\vvv{\uplambda} = 0} = \xi_n(\vv{k})\partial_{k_i}\xi_n(\vv{k})$. \\
%
%
\indent Similarly we find that the second-order derivative of the quasiparticle energy contains, in addition to a contribution determined by the dispersion of the energy bands [the second-order expansion coefficient in Eq.~\eqref{eq:quasi-energy}], also a term governed by the second-order derivative of the fidelity $F_{n\vv{k}}(\vvv{\uplambda})$ with respect to $\lambda_i$ and $\lambda_j$. It can be, 
%
%
similar to the first-order derivative, expressed in terms of derivatives of $\ket{\zeta_n(\vv{k},\vvv{\uplambda})}$. At zero external gravitomagnetic vector potential, the Wilczek-Zee connection of an eigenstate in $\vvv{\uplambda}$ space and that of its complex-conjugated counterpart at opposite momentum differ only by a sign. As a consequence, the second order derivative can be expressed in terms of the quantum metric in the parameter space spanned by the components of the gravitomagnetic vector potential, cf.\ Eq.~\eqref{eq:quantum-metric-boost-deformation-space}, via 
\begin{align}
    \frac{\partial^2 F_{n\vv{k}}(\vvv{\uplambda})}{\partial\lambda_i \partial \lambda_j}\klsr{}_{\vvv{\uplambda} = 0} = -8 g_{ij}^{\ket{\psi_n(\vv{k},\cdot)}}(0)\,.
\end{align}
%
According to the relation provided in Eq.~\eqref{eq:relation-boost-deformed-WZ-and-usual-WZ} between the Wilczek-Zee connections in $\vvv{\uplambda}$ space and momentum space, the quantum metric in $\vvv{\uplambda}$ space can be interpreted as a \enquote{weighted} quantum metric in momentum space. It is proportional to the momentum-space quantum metric in systems with two single-particle energy bands. This quantum-geometric quantity is referred to as the heat quantum metric in Ref.~\cite{bermond2025dichroism} and as the thermal quantum metric in Ref.~\cite{lhachemi2026unifying}. \\
\indent Inspired by the nomenclature in Ref.~\cite{peotta2015superfluidity} used for the superfluid weight, we can divide the contributions to the thermal Meissner stiffness into two contributions $ D_{\mathrm{Q},ij} = D_{\mathrm{Q},ij}^{\mathrm{conv}} + D_{\mathrm{Q},ij}^{\mathrm{geom}}$.
%
%
One contribution depends on derivatives of the energy bands which we call the conventional contribution $D_{\mathrm{Q}}^{\mathrm{conv}}$, and the other contribution depends on the quantum geometry of the eigenstates which we call the geometrical contribution~$D_{\mathrm{Q}}^{\mathrm{geom}}$. The conventional contribution is given by 
 \begin{align}
    &D_{\mathrm{Q},ij}^{\mathrm{conv}} = \frac{1}{V} \sum_{\vv{k} \in \mathrm{BZ}} \sum_{n=1}^{N_\mathrm{B}} \Bigg[ 2n_{\mathrm{F}}'( E_{n\vv{k}}) \xi_n^2(\vv{k}) \partial_{k_i} \xi_n(\vv{k}) \partial_{k_j} \xi_n(\vv{k}) \nonumber \\
    &+ \bigg[\frac{[2n_{\mathrm{F}}(E_{n\vv{k}}) - 1] \xi_n(\vv{k})}{E_{n\vv{k}}} + 1\bigg] \bigg[2\xi_n(\vv{k})\partial_{k_i}\xi_n(\vv{k}) \partial_{k_j}\xi_n(\vv{k}) \nonumber\\
    &+ \xi_n^2(\vv{k}) \partial_{k_i} \partial_{k_j} \xi_n(\vv{k})\bigg] \Bigg] \,,
    \label{eq:conv-contribution}
\end{align}
where the additional term in the second line is due to the second term in the grand potential~\eqref{eq:grand-potential}. The geometrical contribution is given by
\begin{align}
    D_{\mathrm{Q},ij}^{\mathrm{geom}} &= \frac{1}{V} \sum_{\vv{k} \in \mathrm{BZ}} \sum_{n=1}^{N_\mathrm{B}}  \frac{\kle{1 -  2n_{\mathrm{F}}(E_{n\vv{k}})} |\Delta(\vv{k})|^2}{E_{n\vv{k}}} \nonumber \\
    &\times \sum_{m\neq n} \kle{\xi_n(\vv{k}) + \xi_m(\vv{k})}^2 g_{ij,m}^{(n)}(\vv{k}) \,,
    \label{eq:geometrical-contribution}
\end{align}
where $g_{ij,m}^{(n)}(\vv{k}) = \mathrm{Re}\big[\bar{e}_{i,m}^{(n)}(\vv{k}) e_{j,m}^{(n)}(\vv{k})\big]$ is the band-resolved quantum metric \cite{iskin2024cooper}. In the case of a two-band system, the band-resolved quantum metric reduces to 
%
%
the quantum metric of the two-band system in momentum space. 
\section{Wiedemann-Franz-type inequality}
Suppose the band with index $n_0 \in \klg{1,\hdots,N_{\mathrm{B}}}$ is flat with dispersion $\varepsilon_{n_0}(\vv{k}) \approx \mu$. Then, since we assume isolated bands, $E_{n_0\vv{k}} \ll E_{n\vv{k}}$ for all $n \neq n_0$ such that the conventional contribution \eqref{eq:conv-contribution} is zero and only the summand with index $n_0$ survives in the geometrical contribution \eqref{eq:geometrical-contribution}. The dominant contribution to the thermal Meissner stiffness is provided by
\begin{align}
    D_{\mathrm{Q},ij} &= \frac{1}{V} \sum_{\vv{k}\in\mathrm{BZ}} \frac{1 - 2n_{\mathrm{F}}(E_{n_0\vv{k}})}{E_{n_0\vv{k}}} |\Delta(\vv{k})|^2 \sum_{m \neq n_0} \xi_m(\vv{k})^2 g_{ij,m}^{(n_0)}(\vv{k}) \,.
\end{align}
Both the band-resolved quantum metric and the thermal Meissner stiffness are positive semidefinite matrices. Recall results on the superfluid weight $D_{\mathrm{S}}$ in the isolated flat-band limit for TRS preserving gap functions established in Ref.~\cite{buthenhoff2025low}
\begin{align}
    D_{\mathrm{S},ij} &= \frac{4}{V} \sum_{\vv{k}\in\mathrm{BZ}} \frac{1 - 2n_{\mathrm{F}}(E_{n_0\vv{k}})}{E_{n_0\vv{k}}}|\Delta(\vv{k})|^2 g_{ij}^{(n_0)}(\vv{k})\,,
\end{align}
where $g_{ij}^{(n)}(\vv{k}) = \sum_{m\neq n} g_{ij,m}^{(n)}(\vv{k})$. We then find that the thermal Meissner stiffness is sandwiched by the superfluid weight in Löwner sense \cite{Lowner1934}, with prefactors given by the maximal and minimal values of the squared energy offsets over momentum space and all bands $m\neq n_0$. 
%
%
As a consequence, the ratio of determinants of thermal Meissner stiffness and superfluid weight admits the bounds
\begin{align}
     \min_{\vv{k}, m\neq n_0}\bigg[\frac{\xi_m(\vv{k})}{2}\bigg]^{2d} \le \frac{\mathrm{det}(D_{\mathrm{Q}})}{\mathrm{det}(D_{\mathrm{S}})} \le \max_{\vv{k}, m\neq n_0} \bigg[\frac{\xi_m(\vv{k})}{2}\bigg]^{2d},
    \label{eq:bound-final}
\end{align}
which is a Wiedemann-Franz-type inequality for the thermal Meissner stiffness $D_{\mathrm{Q}}$ and electric Meissner stiffness $D_{\mathrm{S}}$ of a superconducting system. Due to the Löwner order between the superfluid weight and the thermal Meissner stiffness, a similar inequality holds for the ratio of $\tr\!\,(D_{\mathrm{Q}})/\tr(D_{\mathrm{S}})$. The only difference is that the exponents appearing inside the minimum and maximum must be replaced by $2$. \\
\indent The bound in Eq.~\eqref{eq:bound-final} has a couple of consequences. First of all, a nonzero geometrical superfluid weight implies a nonzero geometrical thermal Meissner stiffness, and vice versa. Second, we can use the thermal Meissner stiffness to estimate the Berezinskii-Kosterlitz-Thouless (BKT) transition temperature~\cite{berezinskii1971destruction,kosterlitz1973ordering} with the help of the Kosterlitz-Thouless equation~\cite{kosterlitz1972long,huhtinen2022revisiting} in two-dimensional materials via 
\begin{align}
    T_{\mathrm{BKT}} \le \frac{\pi}{2} \max_{\vv{k}, m\neq n_0}\kle{\frac{1}{\xi_m(\vv{k})^{2}}} \sqrt{\det\,\![D_{\mathrm{Q}}(T=0)]} \,.
\end{align}
Moreover, in case the minimum and maximum values of $\xi_m(\vv{k})$ are close to one another, which may be for example the case for two-band systems, both stiffnesses become approximately proportional. Its proportionality coefficient is a material-dependent coefficient set by the typical squared energy offset of the outer bands from the chemical potential. In particular, we expect the ratio to be approximately the same value for different temperatures but the same material. Lastly, under the assumptions made here, it has been shown in Ref.~\cite{peotta2015superfluidity} that the determinant of the superfluid weight admits a lower bound in terms of the Chern number due to the Wirtinger inequality between the quantum metric and the Berry curvature~\cite{wirtinger1936determinantenidentitat,roy2014band,ozawa2021relations,mera2021kahler,mera2022relating}. Since the determinant of the thermal Meissner stiffness admits a lower bound in terms of the determinant of the superfluid weight, we also have a lower bound in terms of the Chern number in the present case. 
\section{Concluding remarks}
We have calculated the thermal Meissner stiffness for fermionic superfluids such as superconductors with isolated bands hosting a time-reversal-symmetry-preserving gap function. In particular, we identified a contribution determined by the quantum geometry, specifically, the quantum metric in the parameter space spanned by the components of the gravitomagnetic vector potential, of the single-particle eigenstates, which can be expressed in terms of Wilczek-Zee connections in momentum space. In the limit of flat bands, we find that the thermal Meissner stiffness admits lower and upper bounds in terms of the superfluid weight, leading to an upper and a lower Wiedemann-Franz-type inequality in which the prefactors depend on the single-particle dispersion of the outer bands of the system. \\
\indent There are multiple directions for extending this work. First, one could generalize this framework to arbitrary (time-reversal-symmetry-breaking) pairing and to non-isolated bands. Second, it would be valuable to evaluate our bounds and geometric contributions in concrete lattice models and flat-band superconductors \cite{kopnin2011high,heikkila2016flat,tian2023evidence}, such as twisted bilayer graphene \cite{cao2018unconventional,cao2018correlated} or other moir\'e superconductors~\cite{balents2020superconductivity,andrei2021marvels,hao2021electric,park2021tunable,xia2025superconductivity,guo2025superconductivity}, in order to identify regimes where the results of this work become experimentally accessible. Since the external gravitomagnetic vector potential is artificial and purely theoretical in nature, we do not expect a \enquote{thermal Meissner effect,} understood as the expulsion of a gravitomagnetic field, to be experimentally measurable. However, the thermal Meissner stiffness itself is measurable and can be determined experimentally by extracting the zero-frequency peak of the thermal conductivity according to Eq.~\eqref{eq:thermal-cond}. The quantum-geometric contribution to the thermal Meissner stiffness can then be verified in a manner analogous to the Meissner stiffness in Ref.~\cite{tian2023evidence}, because the conventional and geometric contributions exhibit distinct dependences on the order parameter. It is however necessary to take into account the impact of possible competing mechanisms for heat transport. Lastly, it would be interesting to investigate higher-order gravitomagnetic responses. We expect these to give rise to higher-order quantum-geometric quantities, such as the quantum Christoffel symbol or the quantum Riemann curvature tensor~\cite{hetenyi2023fluctuations}. \\
\indent Although our analysis is formulated for crystalline systems described by Bloch bands, the underlying idea that the thermal (and charge) responses of superconducting states can be shaped by geometric properties of the quantum states and by their coupling to effective gravitational fields may have broader implications. In particular, superfluid and superconducting phases in neutron star interiors \cite{baym1969superfluidity,sauls1989superfluidity,wood2022superconducting,universe10030104}, which coexist with strong gravitational and rotational fields, might exhibit signatures of geometric contributions to their thermal or rotational response. Indeed, Refs.~\cite{almirante2025superfluid,almirante2026emergence} discuss the superfluid density of neutron star interiors and identify an additional quantum-geometric contribution due to a small relative velocity between superfluid and normal components. Similarly, an extension of our framework to the strongly interacting relativistic regime relevant for astrophyiscal matter might reveal a similar observation. 
\begin{acknowledgments}
We thank Ryota Nakai for valuable discussion and comments on our manuscript. This work was supported by the doctoral scholarship program of the German Academic Scholarship Foundation (M.B.) and by JSPS KAKENHI Grant No.~JP21K03384 (Y.N.).
\end{acknowledgments}

\section*{Data availability}
No data were created or analyzed in this study.

\appendix
\section{General formula for the thermal Meissner stiffness}\label{appendix}
In Ref.~\cite{nakai2022energy}, see also Ref.~\cite{Nakai2023twisted}, it has been shown that the thermal Meissner stiffness admits the general formula
\begin{align}
    D_{\mathrm{Q},ij} = \frac{1}{V} \frac{\mathrm{d}^2F}{\mathrm{d}\lambda_i\mathrm{d}\lambda_j} \klsr{}_{\vvv{\uplambda} = 0} \,,
    \label{eq:tms-definition}
\end{align}
where $F$ is the free energy of the system. In this Appendix, we show the validity of Eq.~\eqref{eq:thermal-meissner-here} given the assumptions that the single-particle Hamiltonian is time-reversal symmetric, and that the gap function is a real-valued function which preserves TRS. We follow the derivation of the Supplemental Material of Ref.~\cite{peotta2015superfluidity}. \\
\indent We need to determine the total derivative of the free energy $F = \Omega + \mu N$ (with fixed particle number $N$) with respect to the components of the external gravitomagnetic vector potential. The total derivative of the free energy with respect to the components of the external gravitomagnetic vector potential is given by
\begin{align}
    \frac{\mathrm{d} F}{\mathrm{d}\lambda_i} = \frac{\mathrm{d}\Omega}{\mathrm{d}\lambda_i} + N \frac{\mathrm{d}\mu}{\mathrm{d}\lambda_i} = \frac{\partial\Omega}{\partial\lambda_i} + \frac{\partial\Omega}{\partial\mu}\frac{\mathrm{d}\mu}{\mathrm{d}\lambda_i} + N\frac{\mathrm{d}\mu}{\mathrm{d}\lambda_i} \,,
\end{align}
where we can ignore the dependence of the gap functions on $\vvv{\uplambda}$ due to TRS. Since, by definition, $N = -\partial\Omega/\partial\mu$, the last two terms cancel and we are left with
\begin{align}
    \frac{\mathrm{d}F}{\mathrm{d}\lambda_i} = \frac{\partial\Omega}{\partial\lambda_i} \,.
\end{align}
Using this result, at $\vvv{\uplambda} = 0$, the second-order total derivative of the free energy with respect to the components of the gravitomagnetic vector potential is given by
\begin{align}
    \frac{\mathrm{d}^2F}{\mathrm{d}\lambda_i\mathrm{d}\lambda_j}\klsr{}_{\vvv{\uplambda} = 0} = \frac{\mathrm{d}}{\mathrm{d}\lambda_j} \frac{\partial\Omega}{\partial\lambda_i}\klsr{}_{\vvv{\uplambda} = 0} = \frac{\partial^2\Omega}{\partial\lambda_i \partial\lambda_j}\klsr{}_{\vvv{\uplambda} = 0} + \frac{\partial^2\Omega}{\partial\lambda_i \partial\mu}\klsr{}_{\vvv{\uplambda} = 0} \frac{\mathrm{d}\mu}{\mathrm{d}\lambda_j}\klsr{}_{\vvv{\uplambda} = 0} \,.
\end{align}
Using $N = -\partial\Omega/\partial\mu$, we further find
\begin{align}
    \frac{\mathrm{d}^2F}{\mathrm{d}\lambda_i\mathrm{d}\lambda_j}\klsr{}_{\vvv{\uplambda} = 0} =\frac{\partial^2\Omega}{\partial\lambda_i \partial\lambda_j}\klsr{}_{\vvv{\uplambda} = 0} - \frac{\partial N}{\partial \lambda_i}\klsr{}_{\vvv{\uplambda} = 0} \frac{\mathrm{d}\mu}{\mathrm{d}\lambda_j}\klsr{}_{\vvv{\uplambda} = 0} \,,
\end{align}
where we note that, when $N = N(\mu(\vvv{\uplambda}),\vvv{\uplambda})$ is held fixed, only the total derivative of $N$ with respect to the components of the gravitomagnetic vector potential is zero. Hence, the thermal Meissner stiffness is determined by two contributions:
\begin{align}
    D_{\mathrm{Q},ij} = \frac{1}{V} \frac{\partial^2\Omega}{\partial \lambda_i \partial \lambda_j}\klsr{}_{\vvv{\uplambda} = 0} - \frac{1}{V} \frac{\partial N}{\partial \lambda_i}\klsr{}_{\vvv{\uplambda} = 0} \frac{\mathrm{d}\mu}{\mathrm{d}\lambda_j}\klsr{}_{\vvv{\uplambda} = 0} \,.
    \label{eq:general-formula}
\end{align}
Suppose the matrix $W_{\vv{k}}(\vvv{\uplambda})$ diagonalizes the BdG Hamiltonian \eqref{eq:bdg-hamiltonian} within the isolated-bands limit. Then, the expectation value of the matrix $\kla{\Phi_{\vv{k}} \otimes \Phi^\dagger_{\vv{k}}}_{\vvv{\uplambda}}$, which also contains the particle number $N(\mu(\vvv{\uplambda}),\vvv{\uplambda}) = \sum_{\vv{k},\alpha} \kla{\phi^\dagger_{\vv{k}\alpha} \phi_{\vv{k}\alpha}}_{\vvv{\uplambda}}$, is given by \cite{peotta2015superfluidity}
\begin{align}
    \kla{\Phi_{\vv{k}} \otimes \Phi^\dagger_{\vv{k}}}_{\vvv{\uplambda}} = W_{\vv{k}}(\vvv{\uplambda}) n_{\mathrm{F}}(E_{\vv{k}}(\vvv{\uplambda})) W^\dagger_{\vv{k}}(\vvv{\uplambda}) \,,
\end{align}
where we defined the diagonal matrix $E_{\vv{k}}(\vvv{\uplambda}) = \mathrm{diag}(E_{+1\vv{k}}(\vvv{\uplambda}),\hdots,E_{+N_{\mathrm{B}}\vv{k}}(\vvv{\uplambda}),E_{-1\vv{k}}(\vvv{\uplambda}),\hdots,,E_{-N_{\mathrm{B}}\vv{k}}(\vvv{\uplambda}))$ with eigenvalues provided in Eq.~\eqref{eq:quasi-energy}. At $\vvv{\uplambda} = 0$, the diagonal matrix reduces to 
\begin{align}
    E_{\vv{k}} = \mathrm{diag}(+E_{1\vv{k}},\hdots,+E_{N_{\mathrm{B}}\vv{k}},-E_{1\vv{k}},\hdots,-E_{N_{\mathrm{B}}\vv{k}}) \,,
\end{align}
where $E_{\vv{k}} \equiv E_{\vv{k}}(0)$. We take the first derivative with respect to $\lambda_i$. According to the chain rule, we obtain
\begin{align}
    \partial_{\lambda_i} \kla{\Phi_{\vv{k}} \otimes \Phi^\dagger_{\vv{k}}}_{\vvv{\uplambda}} &= W_{\vv{k}}(\vvv{\uplambda}) \Big[\partial_{\lambda_i} n_{\mathrm{F}}(E_{\vv{k}}(\vvv{\uplambda})) \nonumber \\
    &+ [W^\dagger_{\vv{k}}(\vvv{\uplambda}) \partial_{\lambda_i} W_{\vv{k}}(\vvv{\uplambda}),n_{\mathrm{F}}(E_{\vv{k}}(\vvv{\uplambda}))]\Big] W_{\vv{k}}^\dagger(\vvv{\uplambda}) \,.
\end{align}
Let $\Gamma_i(\vvv{\uplambda}) = W^\dagger_{\vv{k}}(\vvv{\uplambda}) \partial_{\lambda_i} W_{\vv{k}}(\vvv{\uplambda})$. Then, the diagonal elements of the commutator are given by
\begin{align}
    [\Gamma_i(\vvv{\uplambda}),n_{\mathrm{F}}(E_{\vv{k}}(\vvv{\uplambda}))]_{\alpha\alpha} &= \sum_\beta \big([\Gamma_i(\vvv{\uplambda})]_{\alpha\beta} [n_{\mathrm{F}}(E_{\vv{k}}(\vvv{\uplambda}))]_{\beta\alpha} \nonumber\\
    &-  [n_{\mathrm{F}}(E_{\vv{k}}(\vvv{\uplambda}))]_{\alpha\beta} [\Gamma_i(\vvv{\uplambda})]_{\beta\alpha}\big)\,.
\end{align}
Since $E_{\vv{k}}(\vvv{\uplambda})$ is a diagonal matrix, we find
%
    $[\Gamma_i(\vvv{\uplambda}),n_{\mathrm{F}}(E_{\vv{k}}(\vvv{\uplambda}))]_{\alpha\alpha} 
    = 0$. 
%
On the other hand, the offdiagonal elements of the commutator with $\alpha \neq \alpha'$ can be expressed with the help of the Hellmann-Feynman theorem as
\begin{align}
    [\Gamma_i(\vvv{\uplambda}),n_{\mathrm{F}}(E_{\vv{k}}(\vvv{\uplambda}))]_{\alpha\alpha'} &= \frac{n_{\mathrm{F}}(E_{\alpha\vv{k}}(\vvv{\uplambda})) - n_{\mathrm{F}}(E_{\alpha'\vv{k}}(\vvv{\uplambda}))}{E_{\alpha\vv{k}}(\vvv{\uplambda}) - E_{\alpha'\vv{k}}(\vvv{\uplambda})}\nonumber\\
    &\times [W^\dagger_{\vv{k}}(\vvv{\uplambda}) \partial_{\lambda_i} \mathcal{H}_{\mathrm{BdG}}(\vv{k},\vvv{\uplambda}) W_{\vv{k}}(\vvv{\uplambda})]_{\alpha\alpha'}.
\end{align}
Hence, if we evaluate the first derivative at $\vvv{\uplambda} = 0$, we find with $W_{\vv{k}} \equiv W_{\vv{k}}(0)$
\begin{widetext}
\begin{align}
    \frac{\partial}{\partial\lambda_i}\kla{\Phi_{\vv{k}} \otimes \Phi^\dagger_{\vv{k}}}_{\vvv{\uplambda}} \klsr{}_{\vvv{\uplambda} = 0} &= W_{\vv{k}} \Bigg[\frac{\partial E_{\vv{k}}(\vvv{\uplambda})}{\partial \lambda_i} \klsr{}_{\vvv{\uplambda} = 0} n_{\mathrm{F}}'(E_{\vv{k}}) + \sum_{\alpha\neq \alpha'}\ket{\alpha}\frac{n_{\mathrm{F}}(E_{\alpha\vv{k}}) - n_{\mathrm{F}}(E_{\alpha'\vv{k}})}{E_{\alpha\vv{k}} - E_{\alpha'\vv{k}}} [W^\dagger_{\vv{k}} \partial_{\lambda_i} \mathcal{H}_{\mathrm{BdG}}(\vv{k},\vvv{\uplambda})|_{\vvv{\uplambda}=0} W_{\vv{k}}]_{\alpha\alpha'}\bra{\alpha'} \Bigg] W_{\vv{k}}^\dagger \,,
\end{align}
where $\klg{\ket{\alpha}}_{\alpha = 1,\hdots,2N_{\mathrm{B}}} = \klg{\ket{\sigma n}}_{\sigma=\pm,n = 1,\hdots,N_{\mathrm{B}}}$ denotes the canonical basis of the Hilbert space spanned by the eigenfunctions. Now we insert the heat current operator and $\partial_{\lambda_i} E_{\pm n\vv{k}}(\vvv{\uplambda})|_{\vvv{\uplambda} = 0} = \xi_n(\vv{k})\partial_{k_i}\xi_n(\vv{k})$ to find
\begin{align}
    \frac{\partial}{\partial\lambda_i}\kla{\Phi_{\vv{k}} \otimes \Phi^\dagger_{\vv{k}}}_{\vvv{\uplambda}} \klsr{}_{\vvv{\uplambda} = 0} &= W_{\vv{k}} \Bigg[\matr{\xi(\vv{k})\partial_{k_i}\xi(\vv{k}) & 0\\ 0 & \xi(\vv{k})\partial_{k_i}\xi(\vv{k})} n_{\mathrm{F}}'(E_{\vv{k}}) \nonumber \\
    &- \sum_{\alpha \neq \alpha'}\ket{\alpha}\frac{n_{\mathrm{F}}(E_{\alpha\vv{k}}) - n_{\mathrm{F}}(E_{\alpha'\vv{k}})}{E_{\alpha\vv{k}} - E_{\alpha'\vv{k}}} \kle{W^\dagger_{\vv{k}} \matr{J_i^{\mathrm{Q}}(\vv{k}) & 0\\ 0 & J_i^{\mathrm{Q}}(\vv{k})} W_{\vv{k}}}_{\alpha\alpha'}\bra{\alpha'} \Bigg] W_{\vv{k}}^\dagger \,.
\end{align}    
%
Since the gap function is real valued and preserves TRS, at $\vvv{\uplambda} = 0$, the matrix $R_{n\vv{k}}$ that diagonalizes the ($2 \times 2$) block matrices in Eq.~\eqref{eq:small-bdg-matrix} has the shape
\begin{align}
    R_{n\vv{k}} = \matr{U_{n\vv{k}} & -V_{n\vv{k}}\\ V_{n\vv{k}} & U_{n\vv{k}}}\,,
\end{align}
where 
\begin{align}
    U_{n\vv{k}} &= \sqrt{\frac{1}{2}\klr{1 + \frac{\xi_n(\vv{k})}{E_{n\vv{k}}}}}\,, \qquad
    V_{n\vv{k}} = \mathrm{sgn}(\Delta(\vv{k})) \sqrt{\frac{1}{2}\klr{1 - \frac{\xi_n(\vv{k})}{E_{n\vv{k}}}}} \,.
\end{align}
Thus, we can decompose $W_{\vv{k}}$ as
\begin{align}
    W_{\vv{k}} = \matr{S(\vv{k}) & 0\\ 0 & S(\vv{k})} \matr{U_{\vv{k}} & -V_{\vv{k}}\\ V_{\vv{k}} & U_{\vv{k}}} = \matr{\Tilde{U}_{\vv{k}} & -V_{\vv{k}} \\V_{\vv{k}} & \Tilde{U}_{\vv{k}}} \,.
\end{align}
Let $A_{\vv{k}} = \Tilde{U}_{\vv{k}}^\dagger J_i^{\mathrm{Q}}(\vv{k}) \Tilde{U}_{\vv{k}} + V_{\vv{k}}^\dagger J_i^{\mathrm{Q}}(\vv{k}) V_{\vv{k}}$ and $B_{\vv{k}} = \Tilde{U}_{\vv{k}}^\dagger J_i^{\mathrm{Q}}(\vv{k}) V_{\vv{k}} - V_{\vv{k}}^\dagger J_i^{\mathrm{Q}}(\vv{k}) \Tilde{U}_{\vv{k}}$. Then, we obtain
%
\begin{align}
    \frac{\partial}{\partial\lambda_i}\kla{\Phi_{\vv{k}} \otimes \Phi^\dagger_{\vv{k}}}_{\vvv{\uplambda}} \klsr{}_{\vvv{\uplambda} = 0} &= W_{\vv{k}} \matr{\xi(\vv{k})\partial_{k_i}\xi(\vv{k}) & 0\\ 0 & \xi(\vv{k})\partial_{k_i}\xi(\vv{k})} \matr{n_{\mathrm{F}}'(+E_{+\vv{k}}) & 0\\ 0 & n_{\mathrm{F}}'(-E_{+\vv{k}})} W_{\vv{k}}^\dagger \nonumber \\
    &- \sum_{\sigma n \neq \sigma' n'} W_{\vv{k}}\ket{\sigma n} \bra{\sigma' n'}W_{\vv{k}}^\dagger \matr{\frac{n_{\mathrm{F}}(E_{+n\vv{k}}) - n_{\mathrm{F}}(E_{+n'\vv{k}})}{E_{+n\vv{k}} - E_{+n'\vv{k}}} [A_{\vv{k}}]_{nn'} & - \frac{n_{\mathrm{F}}(E_{+n\vv{k}}) + n_{\mathrm{F}}(E_{+n'\vv{k}}) - 1}{-E_{+n\vv{k}} - E_{+n'\vv{k}}} [B_{\vv{k}}]_{nn'} \\ \frac{n_{\mathrm{F}}(E_{+n\vv{k}}) + n_{\mathrm{F}}(E_{+n'\vv{k}}) - 1}{E_{+n\vv{k}} + E_{+n'\vv{k}}}[B_{\vv{k}}]_{nn'} & \frac{n_{\mathrm{F}}(E_{+n\vv{k}}) - n_{\mathrm{F}}(E_{+n'\vv{k}})}{E_{+n\vv{k}} - E_{+n'\vv{k}}} [A_{\vv{k}}]_{nn'}}_{\sigma\sigma'}\,.
\end{align}    
\end{widetext}
Note that $n_{\mathrm{F}}'(+E) = n_{\mathrm{F}}'(-E)$. We define the symmetry operator 
\begin{align}
    \tau_y = \matr{0 & -i\\ i & 0}
\end{align}
acting in particle-hole space. Then $\tau_y W_{\vv{k}} \tau_y = W_{\vv{k}}$, and we find
\begin{align}
    \frac{\partial}{\partial\lambda_i}\kla{\Phi_{\vv{k}} \otimes \Phi^\dagger_{\vv{k}}}_{\vvv{\uplambda}} \klsr{}_{\vvv{\uplambda} = 0} = \tau_y \frac{\partial}{\partial\lambda_i}\kla{\Phi_{\vv{k}} \otimes \Phi^\dagger_{\vv{k}}}_{\vvv{\uplambda}} \klsr{}_{\vvv{\uplambda} = 0} \tau_y \,. 
    \label{eq:final-result-end-matter}
\end{align}
The diagonal elements give rise to the equality\\\\
\begin{align}
    \frac{\partial}{\partial{\lambda_i}} \kla{\phi^\dagger_{\vv{k}\alpha} \phi_{ \vv{k}\alpha}}_{\vvv{\uplambda}} \klsr{}_{\vvv{\uplambda} = 0} 
    &= -\frac{\partial}{\partial \lambda_i} \kla{\phi^\dagger_{-\vv{k}\alpha} \phi_{-\vv{k}\alpha}}_{\vvv{\uplambda}} \klsr{}_{\vvv{\uplambda} = 0} \,.
\end{align}
Using this result, after the substitution $\vv{k} \to -\vv{k}$, the first partial derivative of the particle number with respect to the components of the external gravitomagnetic vector potential becomes
\begin{align}
    \frac{\partial N}{\partial \lambda_i}\klsr{}_{\vvv{\uplambda} = 0} = 0\,,
\end{align}
as required.

\bibliography{apssamp}

\end{document}